\documentclass[conference]{IEEEtran}
\IEEEoverridecommandlockouts
\usepackage{cite}
\usepackage{amsmath,amssymb,amsfonts}
\usepackage{algorithmic}
\usepackage{graphicx}
\usepackage{textcomp}
\usepackage{xcolor}
\usepackage{amsthm}
\usepackage[abs]{overpic}

\theoremstyle{noparens}

\graphicspath{{./figures/} }
\DeclareGraphicsExtensions{.pdf,.jpeg,.png}

\definecolor{mygreen}{RGB}{14,110,14}

\def\BibTeX{{\rm B\kern-.05em{\sc i\kern-.025em b}\kern-.08em
    T\kern-.1667em\lower.7ex\hbox{E}\kern-.125emX}}
\begin{document}

\title{Determined Multichannel Blind Source Separation with Clustered Source Model}

\author{\IEEEauthorblockN{Jianyu Wang\IEEEauthorrefmark{1},
shanzheng Guan\IEEEauthorrefmark{1}}
\IEEEauthorblockA{\IEEEauthorrefmark{1}Northwestern Polytechnical
University, China.
    \{alexwang96, gshanzheng, \}@mail.nwpu.edu.cn,}
}

\maketitle

\begin{abstract}

The independent low-rank matrix analysis (ILRMA) method stands out as a
prominent technique for multichannel blind audio source separation. It
leverages nonnegative matrix factorization (NMF) and nonnegative canonical
polyadic decomposition (NCPD) to model source parameters. While it
effectively captures the low-rank structure of sources, the NMF model
overlooks inter-channel dependencies. On the other hand, NCPD preserves
intrinsic structure but lacks interpretable latent factors, making it
challenging to incorporate prior information as constraints. To address
these limitations, we introduce a clustered source model based on
nonnegative block-term decomposition (NBTD). This model defines blocks as
outer products of vectors (clusters) and matrices (for spectral structure
modeling), offering interpretable latent vectors. Moreover, it enables
straightforward integration of orthogonality constraints to ensure
independence among source images. Experimental results demonstrate that our
proposed method outperforms ILRMA and its extensions in anechoic conditions
and surpasses the original ILRMA in simulated reverberant environments.
\end{abstract}

\begin{IEEEkeywords}
Independent low rank matrix analysis, multichannel blind audio source separation, block term decomposition.
\end{IEEEkeywords}

\section{Introduction}

In real-world applications, the observation of a source signal of interest
frequently suffers from interference. Therefore, it is essential to use
signal processing techniques to extract latent sources from observed mixtures
by multiple microphones \cite{huang2006acoustic, belouchrani1997blind}.
Generally, there are two different paradigms for source separation. One is
beamforming, which extracts the components from mixed signals using direction
information with spatial filtering techniques \cite{Benesty2008Microphone,
benesty2017fundamentals, lee2011beamspace}. They usually assume that the
geometry of array and the incidence angle of each source are known. Another
paradigm is to perform multichannel blind audio source separation (MBASS)
based on independent component analysis (ICA) \cite{comon1994independent},
which exploits the statistical independence of source signals. This study
focuses on the latter paradigm.

Generally, MBASS is conducted in the short-time Fourier transform (STFT)
domain to address convolutive mixing. However, a significant challenge arises
due to inner permutations, which can greatly affect separation performance.
To tackle this issue, independent vector analysis (IVA)
\cite{kim2006independent} was then adopted, and the majorization-minimization
(MM) principle \cite{sun2016majorization} was introduced to derive fast
update rules for IVA \cite{ono2011stable}. Despite its effectiveness in
handling permutations, IVA-based methods often overlook spectral structure
information. To incorporate spectral structure, nonnegative matrix
factorization (NMF) \cite{lee2000algorithms} was extended to multichannel
cases for MBASS \cite{ozerov2009multichannel, duong2010under,
sawada2013multichannel}. However, Multichannel NMF methods are
computationally demanding and sensitive to parameter initialization. In
response to these challenges, the independent low-rank matrix analysis
(ILRMA) method was devised \cite{kitamura2016determined}.  ILRMA enforces an
interpretable low-rank structure constraint on the spectrogram and employs a
rank-one relaxation for the spatial model.

To further improve performance, efforts have been directed towards enhancing
the source model by generalizing the distribution of source signals and
achieving initialization-robust performance \cite{mogami2017independent,
kitamura2018generalized, ikeshita2018independent, mogami2018independent,
mogami2019independent}. Additionally, research has explored the integration
of prior constraints into parameters associated with the source model to
enhance MBASS performance \cite{mitsui2017blind, wang2021minimum}. However,
these NMF-based methods are typically inadequate to capture inter-channel
dependencies and higher-order structures inherent in multichannel data. While
Kitamura \cite{kitamura2016determined} also discussed a CPD-based source
model, its spectral distinctiveness remains somewhat deficient.

In this paper, we propose a clustered source model for determined blind
source separation. Utilizing nonnegative block-term decomposition (NBTD), the
source model parameters are expressed as a summation of several components,
each being the outer product of a vector and a matrix. By applying
orthogonality constraints to the latent vectors of this decomposition, they
gain a clear interpretation, revealing distinct clusters of sources.

\section{Signal model and problem formulation}

Let $M$ and $N$ be the number of microphones and sources, respectively. The
observed signal in the STFT domain can be expressed as:
\begin{align}\label{Mix}
  \mathbf{x}_{ij} = \sum_{n=1}^{N} \mathbf{c}_{nij} = \mathbf{A}_i \mathbf{s}_{ij}, \quad 1 \leq i \leq I, \quad 1 \leq j \leq J,
\end{align}
where the subscripts $i$ and $j$ denote, respectively, the frequency and
frame indices, $I$ and $J$ denote, the total number of STFT bins and time
frames, $\mathbf{x}_{ij} = \left[ \begin{array}{ccc} x_{1ij} & \dots &
x_{Mij}
\end{array} \right]^\intercal  \in \mathbb{C}^M$ and $\mathbf{s}_{ij} =
\left[ \begin{array}{ccc} s_{1ij} & \dots & s_{Nij} \end{array}
\right]^\intercal \in \mathbb{C}^N$ are, respectively, the vectors consisting
of the STFTs of the observation and source signals, respectively,
$\mathbf{c}_{nij} =  \left[
\begin{array}{ccc} c_{nij1} & \dots & c_{nijM} \end{array} \right]^\intercal
\in \mathbb{C}^M$ represents the image of the $n$th source, and $\mathbf{A}_i = \left[
\begin{array}{ccc} \mathbf{a}_{1i} & \dots & \mathbf{a}_{Ni} \end{array}
\right] \in \mathbb{C}^{M\times N}$ is called the mixing matrix. All the signals are assumed to have zero mean.

With the signal model given in (\ref{Mix}), the problem of blind source separation (BSS) becomes one
of identifying a demixing matrix $\mathbf{D}_i$ such that
\begin{align}\label{Separate}
  \mathbf{y}_{ij} = \mathbf{D}_i \mathbf{x}_{ij}, \quad 1 \leq i \leq I, \quad 1 \leq j \leq J,
\end{align}
where $\mathbf{y}_{ij} = \left[ \begin{array}{ccc} y_{1ij} & \dots & y_{Nij} \end{array}\right]^\intercal \in \mathbb{C}^N$
denotes an estimate of the source signal $\mathbf{s}_{ij}$, and
$\mathbf{D}_i = \left[ \begin{array}{ccc} \mathbf{d}_{1i} & \dots & \mathbf{d}_{Mi} \end{array}\right] \in \mathbb{C}^{N\times M}$.
The difficulty of this identifiying depends on many factors, e.g., number of sources, number of sensors, the nature of the
source signals, the property of the mixing system. In this work, we focus on the case where the number of souces is equal
to the number of microphone sensors, i.e.,  $M=N$ and assume that the source signals are mutually independent and their
distributions are stationary. Given these conditions, the problem of MBSS can be solved using only the second-order statistics \cite{belouchrani1997blind}.

From (\ref{Mix}), the covariance matrix of mixtures is $\mathbf{X}_{ij} = \mathbb{E}\left[ \mathbf{x}_{ij} \mathbf{x}_{ij}^{\mathsf{H}} \right]$.
Using the signal model given in \eqref{Mix}, we obtain
\begin{align}\label{SOS2}
  \mathbf{X}_{ij} = \mathbf{A}_i \mathbb{E}\left[  \mathbf{s}_{ij} \mathbf{s}_{ij}^{\mathsf{H}} \right] \mathbf{A}_i^{\mathsf{H}}.
\end{align}
Under the assumption of statistical independence {among} sources, the covariance matrix of sources
$\boldsymbol{\Lambda}_{ij} = \mathbb{E}\left[  \mathbf{s}_{ij} \mathbf{s}_{ij}^{\mathsf{H}} \right]$ should be a diagonal matrix, which can be represented as $\boldsymbol{\Lambda}_{ij} = \mbox{Diag}\left( \begin{array}{cccc} \lambda_{1ij} & \lambda_{2ij} & \cdots & \lambda_{Nij} \end{array} \right)$.
Let us stack all the non-zeros parameters in $\boldsymbol{\Lambda}_{ij}$ to form a large matrix, i.e., 
\begin{align}
\label{Sourcemodel}
\boldsymbol{\lambda} = \left[\begin{array}{ccccccc} \boldsymbol{\lambda}_{1} & \boldsymbol{\lambda}_{2} & \cdots & \boldsymbol{\lambda}_{n} & \cdots & \boldsymbol{\lambda}_{N} \end{array}\right],
\end{align}
where $\boldsymbol{\lambda}_{n}\in\mathbb{R}^{I\times J}$ is a matrix consisting of all the parameters associated with the $n$th source, and $\boldsymbol{\lambda}$
is a matrix of size $N\times I\times J$, i.e., $\boldsymbol{\lambda}\in\mathbb{R}^{N\times I\times J}$, which encompass all the parameters of the source model,

In ILRMA, the NMF tool is used to decompose the matrix $\boldsymbol{\lambda}_{n}$, $n = 1,\cdots, N$, into the following form:
\begin{align}\label{SourceModelILRMA}
  \boldsymbol{\lambda}_n & = \mathbf{T}_n \mathbf{V}_n, 
\end{align}
where $\mathbf{T}_n$ is a basis matrix of size $I\times K$ with $K$ being the number of bases, and $\mathbf{V}_n$ is called the activation matrix, whose dimension is $K \times J $. This
decomposition based source modeling is referred to as the NMF source model.

Another way to model the source model parameters is through a rank-one-tensor decomposition of the matrix $\boldsymbol{\lambda}$, leading to the
non-negative CPD (NCPD) based source model. In NCPD, the matrix $\boldsymbol{\lambda}$ is expressed as
\begin{align}\label{SourceModelILRMAcpd}
  \boldsymbol{\lambda} = & \sum_{k=1}^{K} \mathbf{z}_k \circ \tilde{\mathbf{t}}_k \circ \tilde{\mathbf{v}}_k,
\end{align}
where
\begin{align}
  \mathbf{z}_k = & \left[ \begin{array}{cccc} z_{1k} & z_{2k} & \cdots & z_{Nk} \end{array} \right]^\intercal,  \nonumber \\
  \tilde{\mathbf{t}}_k = & \left[ \begin{array}{cccc} \tilde{t}_{1k} & \tilde{t}_{2k} & \cdots & \tilde{t}_{Ik} \end{array} \right]^\intercal, \nonumber \\
  \tilde{\mathbf{v}}_k = & \left[ \begin{array}{cccc} \tilde{v}_{1k} & \tilde{v}_{2k} & \cdots & \tilde{v}_{Jk} \end{array} \right]^\intercal. \nonumber
\end{align}
and  $\circ$ denotes the outer product.

\subsection{Proposed source model}

\begin{figure}[t]
\begin{minipage}[b]{1\linewidth}
  \centering
  \begin{overpic}[width=8cm]{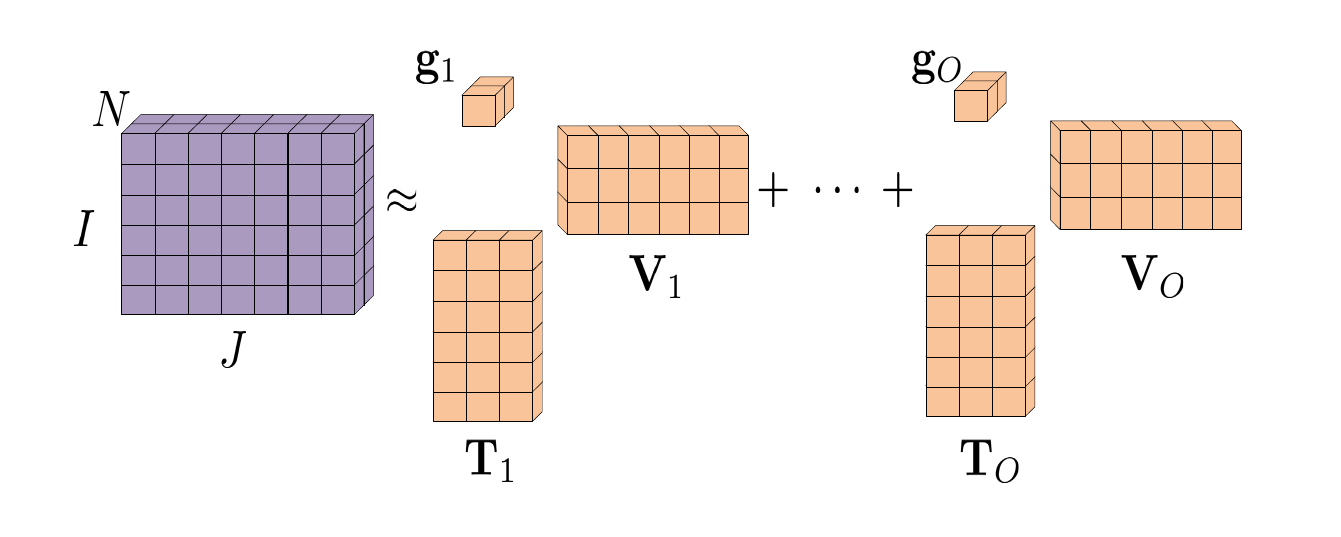}
   \put(78,58){\normalsize $\otimes$}
   \put(162,58){\normalsize $\otimes$}
   \end{overpic}
\end{minipage}
\caption{Illustration of block term decomposition.}
\label{figMVNTR}
\vspace{-0.5cm}
\end{figure}

While the NCPD method adeptly captures the intricate nature of multi-channel data by
expressing the source parameter tensor as a summation of a series of rank-one tensors, it is essential
to acknowledge that multi-channel signals often exhibit a more nuanced structure
in the time-frequency domain. This nuanced structure can be better modeled with the so-called
nonnegative block term decomposition (NBTD), which emerges as a potent tensor
factorization model tailor-made to precisely capture localized and recurring patterns
within the time-frequency representation of speech signals \cite{de2008decompositions}. The
decomposition process is illustrated in Fig.~\ref{figMVNTR} and can be mathematically written as follows:
{
\begin{align}\label{NBTD}
  \boldsymbol{\lambda} & = \sum_{o=1}^O \left( \mathbf{T}_o \mathbf{V}_o^\intercal \right) \circ \mathbf{g}_o, \\
  \mathbf{T}_o & =
  \begin{bmatrix}
  t_{o11} & t_{o12} & \cdots & t_{o1K} \\
  t_{o21} & t_{o22} & \cdots & t_{o2K} \\
  \vdots  & \vdots  & \ddots & \vdots  \\
  t_{oI1} & t_{oI2} & \cdots & t_{oIK}
  \end{bmatrix}, \nonumber \\
  \mathbf{V}_o & =
  \begin{bmatrix}
  v_{o11} & v_{o12} & \cdots & v_{o1J} \\
  v_{o21} & v_{o22} & \cdots & v_{o2J} \\
  \vdots  & \vdots  & \ddots & \vdots  \\
  v_{oK1} & v_{oK2} & \cdots & v_{oKJ}
  \end{bmatrix}, \nonumber
\end{align}
\begin{align}
  \mathbf{g}_o & = \left[ \begin{array}{cccc} g_{1o} & g_{2o} & \cdots & g_{No} \end{array} \right]^\intercal,  \nonumber
\end{align}
where $\boldsymbol{\lambda} \in \mathbb{R}^{N\times I\times J}$, $\mathbf{T}_o \in \mathbb{R}^{I\times L_o}$,
$\mathbf{V}_o \in \mathbb{R}^{J\times L_o}$, and
$\mathbf{G} = \left[ \begin{array}{cccc} \mathbf{g}_{1}&\mathbf{g}_{2}&\dots&\mathbf{g}_{O} \end{array} \right] \in \mathbb{R}^{N\times O}$ are
all non-negative matrices, and $\circ$ denotes the outer product. It is easy to check that in this decomposition
each elements of $\boldsymbol{\lambda}$ is presented as $\lambda_{nij} = \sum_o g_{no} \sum_k t_{oik} v_{okj}$. This decomposition based source
modeling is referred to as the NBTD source model.
}

Let us introduce two matrices:
\begin{align}
  \overset{\circ}{\boldsymbol{\Lambda}}_{ij} & =
\begin{bmatrix}
  \sum_k t_{1ik} v_{1kj} & \dots & 0 \\
  \vdots & \ddots & \vdots \\
  0 & \dots & \sum_k t_{Oik}v_{Okj}
\end{bmatrix}\in\mathbb{R}_{\geq 0}^{O\times O}, \\
   \mathbf{U} & =
\begin{bmatrix}
  g_{11}^{\frac{1}{2}} & g_{12}^{\frac{1}{2}} & \dots & g_{1O}^{\frac{1}{2}} \\
  g_{21}^{\frac{1}{2}} & g_{22}^{\frac{1}{2}} & \dots & g_{2O}^{\frac{1}{2}} \\
  \vdots & \vdots & \ddots & \vdots \\
  g_{N1}^{\frac{1}{2}} & g_{N2}^{\frac{1}{2}} & \dots & g_{NO}^{\frac{1}{2}}
\end{bmatrix}\in\mathbb{R}_{\geq 0}^{N\times O},
\end{align} 
where the matrix $\mathbf{U}$ satisfies the orthogonal constraint.
This constraint not only relates the NBTD based source model to the k-means clustering of spectral components, but also allows the formula \eqref{NBTD} to be expressed in the form of a diagonal matrix under each time-frequency bins:
\begin{align}
  \boldsymbol{\Lambda}_{ij} = \mathbf{U} \overset{\circ}{\boldsymbol{\Lambda}}_{ij} \mathbf{U}^\intercal.
\end{align}

Following the orthogonal constraints on matrix $\mathbf{U}$, we have $\mathbf{U} \mathbf{U}^\intercal = \mathbf{I}_N$. It can also be expressed as:
\begin{align}\label{ClusterG}
  \sum_o g_{no} = 1. \quad \forall \quad n = 1,\dots,N,
\end{align}
It indicates the clusters of sources of source model in cILRMA which fully utilize the spectral structure.
Then, We build the generative model of covariance matrix for cILRMA as:
\vspace{-0.1cm}
\begin{align}\label{oILRMAGenerativeM}
  \hat{\mathbf{X}}_{ij} & = \sum_n \mathbf{a}_{in} \mathbf{a}_{in}^{\mathsf{H}} \left( \sum_o g_{no} \left( \sum_k t_{oik} v_{okj} \right) \right) \nonumber \\[-7bp]
  & = \mathbf{A}_i \mathbf{U} \overset{\circ}{\boldsymbol{\Lambda}}_{ij} \mathbf{U}^\intercal \mathbf{A}_i^{\mathsf{H}}.
\end{align}

Generally, it is assumed that mixed signal in each STFT bin follow a complex Gaussian distribution, i.e.,
\begin{align}
  \mathbf{x}_{ij} \sim \mathcal{N}_{\mathbb{C}} \left( \mathbf{x}_{ij} \bigg| \mathbf{0}, \hat{\mathbf{X}}_{ij} \right).
\end{align}
The maximum likelihood cost function for estimating the model parameters is then written as
\vspace{-0.1cm}
\begin{align}\label{OFoILRMA}
\vspace{-5cm}
  \mathcal{L} =
  & \sum_{ij} \! \Big[ \! \mbox{tr} \! \left( \! \mathbf{D}_i^{\!-\!1} \mathbf{y}_{ij} \mathbf{y}_{ij}^{\mathsf{H}} \! \left( \! \mathbf{D}_i^{\mathsf{H}} \! \right)^{\!-\!1} \! \mathbf{D}_i^{\mathsf{H}} \! \left(\mathbf{U}^\intercal\!\right)^{\!-\!1} ({\overset{\circ}{\boldsymbol{\Lambda}}_{ij}})^{\!-\!1} \mathbf{U}^{\!-\!1} \mathbf{D}_i \! \right) \! \nonumber \\[-7bp]
  & + \log{\left( \det \mathbf{A}_i \right) \left( \det \mathbf{U}\overset{\circ}{\boldsymbol{\Lambda}}_{ij}\mathbf{U}^\intercal \right) \left( \det \mathbf{A}_i^{\mathsf{H}} \right)} \Big] \nonumber \\[-2bp]
  & + \sigma \mbox{tr} \left( \mathbf{U} \mathbf{U}^\intercal - \mathbf{I}_N \right),
\end{align}
where $\sigma$ denotes the Lagrange multiplier with respect to $\mathbf{U}$.
Then the problem of cILRMA is converted into one of estimating the source model related parameters $t_{oik},v_{okj},g_{no}$ and spatial model related parameters $\mathbf{D}_i$.

\section{Parameters Optimization}
In this section, we derive the update rules for source model related parameters using the objective function given in \eqref{OFoILRMA}.
For  $\overset{\circ}{\boldsymbol{\Lambda}}_{ij}$, the the objective function can be expressed as
\begin{align}\label{EstimateLambda1}
  \!\!\!\!\mathcal{L}(\!\overset{\circ}{\boldsymbol{\Lambda}}_{ij}\!)\! = \! \mbox{tr} \! \left( \! \mathbf{y}_{ij}\mathbf{y}_{ij}^{\mathsf{H}}\left( \! \mathbf{U}^\intercal \! \right)^{\!\dag\!} \!\! (\!\overset{\circ}{\boldsymbol{\Lambda}}_{ij}\!)^{\!-1\!}\mathbf{U}^{\!\dag\!} \! \right) \!\! + \! \log\det\!\left( \! \mathbf{U} \overset{\circ}{\boldsymbol{\Lambda}}_{ij} \mathbf{U}^\intercal \! \right) \!\!.\!
\end{align}
Since $\mathbf{U}$ is a unitary orthogonal matrix, \eqref{EstimateLambda1} can be further expressed as
\begin{align}\label{EstimateLambda2}
  & \mathcal{L}(\!\overset{\circ}{\boldsymbol{\Lambda}}_{ij}\!)\! = \mbox{tr}\left( \mathbf{U}^\intercal \mathbf{y}_{ij}\mathbf{y}_{ij}^{\mathsf{H}}\mathbf{U}(\overset{\circ}{\boldsymbol{\Lambda}}_{ij})^{-1} \right) \!\! + \!\! \log\det\left( \mathbf{U} \overset{\circ}{\boldsymbol{\Lambda}}_{ij} \mathbf{U}^\intercal \right)\!,\! \nonumber \\
  & \! = \!\! \sum_{ij} \!\! \Bigg[ \!\! \sum_{o} \!\! \frac{\sum_ng_{no}|y_{nij}|^2}{\sum_k t_{oik}v_{okj}} \! + \!\! \sum_n \! \log \! \sum_o g_{no} \!\! \left( \! \sum_{k} \! t_{oik}v_{okj} \!\! \right) \!\! \Bigg] \! . \!
\end{align}

Direction optimization of $\mathcal{L}(\!\overset{\circ}{\boldsymbol{\Lambda}}_{ij}\!)$ with respect to $\overset{\circ}{\boldsymbol{\Lambda}}_{ij}$ is rather difficult.
To circumvent this, we introduce the Jensen's inequality and the tangent line inequality to obtain the following auxiliary function for \eqref{EstimateLambda2}:
\begin{align}\label{Aux1}
  \mathcal{Q}(\overset{\circ}{\boldsymbol{\Lambda}}_{ij}) = & \sum_{ij} \Bigg[ \sum_{ok} \frac{\sum_ng_{no}|y_{nij}|^2}{t_{oik}v_{okj}} \alpha_{okij}^2 \nonumber \\[-5bp]
  & + \frac{1}{\beta_{oij}} \Bigg( \sum_k t_{oik}v_{okj} - \beta_{oij} \Bigg) + \log\beta_{oij} \Bigg],
\end{align}
where $\alpha_{okij}$ and $\beta_{oij}$ are two auxiliary variables and the equality holds if and only if $\alpha_{okij} = \frac{t_{oik}v_{okj}}{\sum_k t_{oik}v_{okj}}$, $\beta_{oij} = \sum_k t_{oik}v_{okj}$.

Identifying partial derivatives of $\mathcal{Q}(\overset{\circ}{\boldsymbol{\Lambda}}_{ij})$ with respect to $\partial t_{oik}$ and $\partial v_{okj}$
and forcing the results to zero, we obtain:
\begin{align}
  t_{oik} \leftarrow t_{oik} \sqrt{ \frac{ \sum_{nj} |y_{nij}|^2 g_{no} v_{okj} \left( \sum_{k^\prime} t_{oik^\prime} v_{ok^\prime j} \right)^{-2} }{ \sum_{j} v_{okj} \left( \sum_{k^\prime} t_{oik^\prime} v_{ok^\prime j} \right)^{-1} } }, \\
  v_{okj} \leftarrow v_{okj} \sqrt{ \frac{ \sum_{ni} |y_{nij}|^2 g_{no} t_{oik} \left( \sum_{k^\prime} t_{oik^\prime} v_{ok^\prime j} \right)^{-2} }{ \sum_{i} t_{oik} \left( \sum_{k^\prime} t_{oik^\prime} v_{ok^\prime j} \right)^{-1} } }.
\end{align}

Similarly, the maximum likelihood function with respect to $\mathbf{U}$ can be expressed as
\begin{align}\label{EstimateSWM}
  \mathcal{L}(\mathbf{U}) = & \mbox{tr}\left( \mathbf{y}_{ij}\mathbf{y}_{ij}^{\mathsf{H}}\left(\mathbf{U}^\intercal\right)^{-1}(\overset{\circ}{\boldsymbol{\Lambda}}_{ij})^{-1}\mathbf{U}^{-1} \right) \nonumber \\[-5bp]
  & + \log\det\left( \mathbf{U} \overset{\circ}{\boldsymbol{\Lambda}}_{ij} \mathbf{U}^\intercal \right) + \sigma \mbox{tr}\left( \mathbf{U}\mathbf{U}^\intercal - \mathbf{I}_N \right).
\end{align}
Following the previous analysis, one can obtain the auxiliary function for \eqref{EstimateSWM} with respect to $\mathbf{G}$:
\begin{align}\label{Aux2}
  & \mathcal{Q}(\mathbf{G}) \! = \! \sum_{ijn} \!\! \left[ \! \sum_{o} \!\! \frac{|y_{nij}|^2}{g_{no}\!\sum_{k}\!t_{oik}v_{okj}}\hat\alpha_{onij}^2 \! \right] \!\! + \! \sigma \!\! \left( \! \sum_{no} \! g_{no} \! - \! N \!\! \right) \nonumber \\[-3bp]
  & \! + \! \sum_{ijn} \!\! \left[ \! \frac{1}{\hat\beta_{nij}} \! \bigg( \!\! \sum_{o} g_{no} \! \sum_k t_{oik}v_{okj} \! - \! \hat\beta_{nij} \!\! \bigg) \!\! + \! \log\hat\beta_{nij} \!\! \right] \!,\!
\end{align}
where $\hat\alpha_{onij}$ and $\hat\beta_{nij}$ are two auxiliary variables, and the equality satisfies if and only if $\hat{\alpha}_{onij} = \frac{g_{no}\sum_kt_{oik}v_{okj}}{\sum_og_{no}\sum_kt_{oik}v_{okj}}$, and $\hat\beta_{nij} = \sum_o g_{no} \sum_k t_{oik}v_{okj}$.
Identifying the partial derivative of\eqref{Aux2} with respect to $g_{no}$ and forcing the result to zero gives the following update rule:
\begin{align}
  g_{no} \! \leftarrow \! g_{no} \! \sqrt{ \! \frac{ \! \sum_{ij} \! |y_{nij}|^2 \! \left( \! \sum_k  \!t_{oik} v_{okj} \! \right) \! \left( \! \sum_{ok} g_{no} \! t_{oik}v_{okj} \! \right)^{\!-2\!} }{ \sum_{ij}\! \left( \sum_k t_{oik} v_{okj}\! \right) \!\left(\! \sum_{ok} g_{no} t_{oik}v_{okj} \!\right)^{\!-1\!} \! + \! \sigma } } \!.\!
\end{align}

The update rules of the demixing matrix $\mathbf{D}_i$ in cILRMA is similar to that in AuxIVA \cite{ono2011stable}, which are as following:
\begin{align}
  & \mathbf{O}_{ni} = \frac{1}{J} \sum_j \frac{1}{\sum_0 g_{no} \sum_{k} u_{oik} v_{okj}} \mathbf{x}_{ij} \mathbf{x}^{\mathsf{H}}_{ij}, \\
  & \mathbf{d}_{ni} \leftarrow \left[ \mathbf{D}_i \mathbf{O}_{ni} \right]^{-1} \mathbf{e}_n, \\
  & \mathbf{d}_{ni} \leftarrow \mathbf{d}_{ni} \left[ \mathbf{d}_{ni}^{\mathsf{H}} \mathbf{O}_{ni} \mathbf{d}_{ni} \right]^{-\frac{1}{2}}.
\end{align}
where $\mathbf{O}_{ni}$ denotes the auxiliary variable, and $\mathbf{e}_n$ denotes the $n$th column vector of the identity matrix of size $N\times N$.

\begin{figure}[t]
%
\begin{minipage}[b]{.46\linewidth}
  \centering
  \centerline{\includegraphics[width=4.2cm]{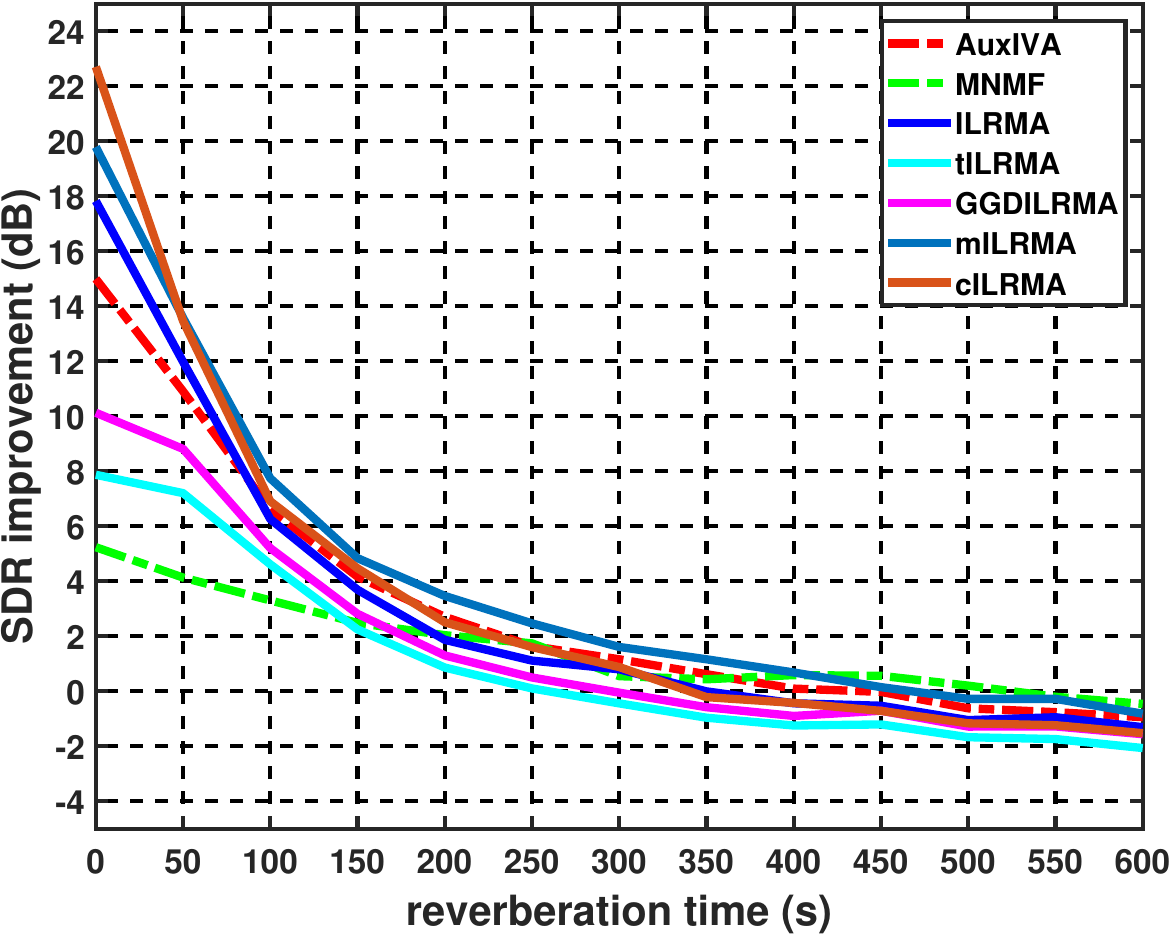}}
  \centerline{(a) female+female}\medskip
\end{minipage}
\begin{minipage}[b]{.55\linewidth}
  \centering
  \centerline{\includegraphics[width=4.2cm]{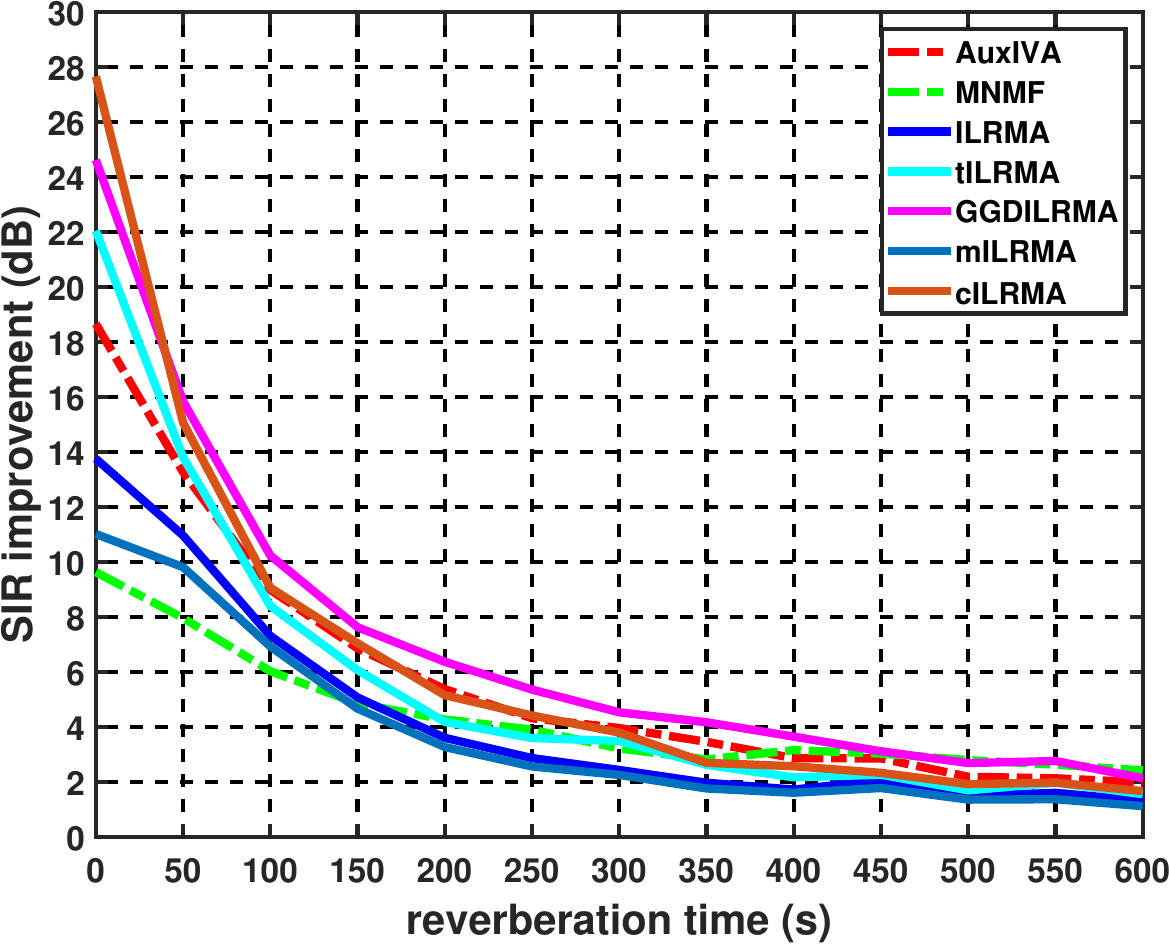}}
  \centerline{(b) female+female}\medskip
\end{minipage}
\vspace{-0.5cm}

\begin{minipage}[b]{0.46\linewidth}
  \centering
  \centerline{\includegraphics[width=4.2cm]{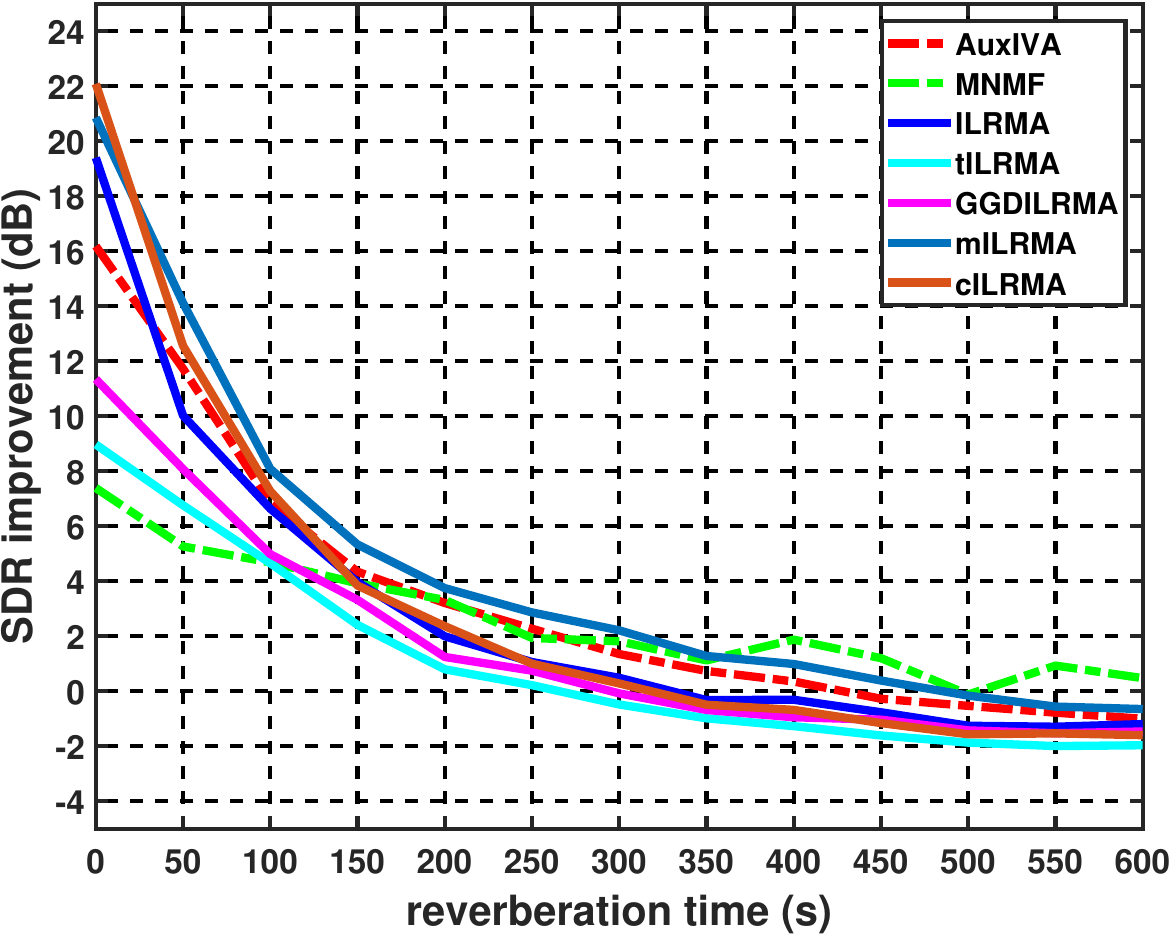}}
  \centerline{(c) male+male}\medskip
\end{minipage}
\begin{minipage}[b]{.55\linewidth}
  \centering
  \centerline{\includegraphics[width=4.2cm]{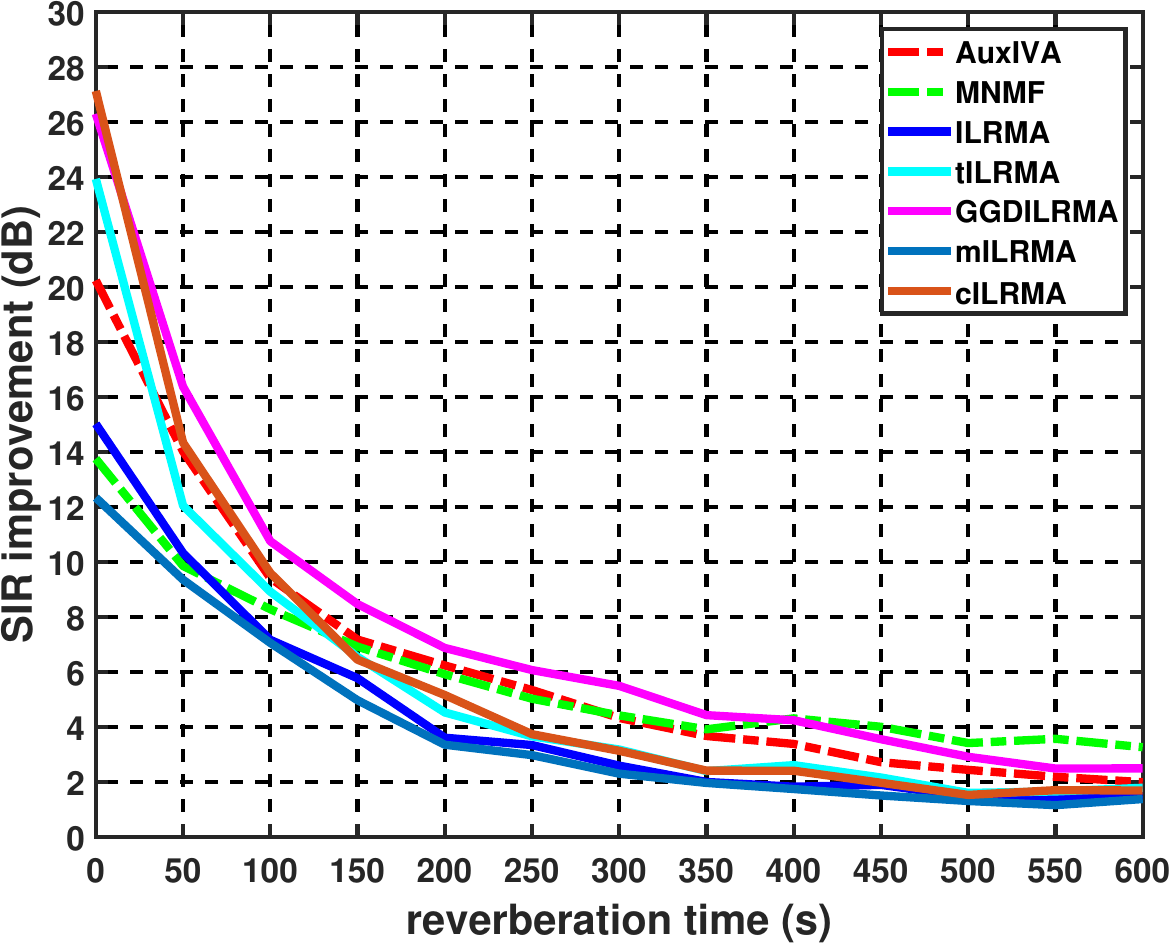}}
  \centerline{(d) male+male}\medskip
\end{minipage}
\vspace{-0.5cm}

\begin{minipage}[b]{.46\linewidth}
  \centering
  \centerline{\includegraphics[width=4.2cm]{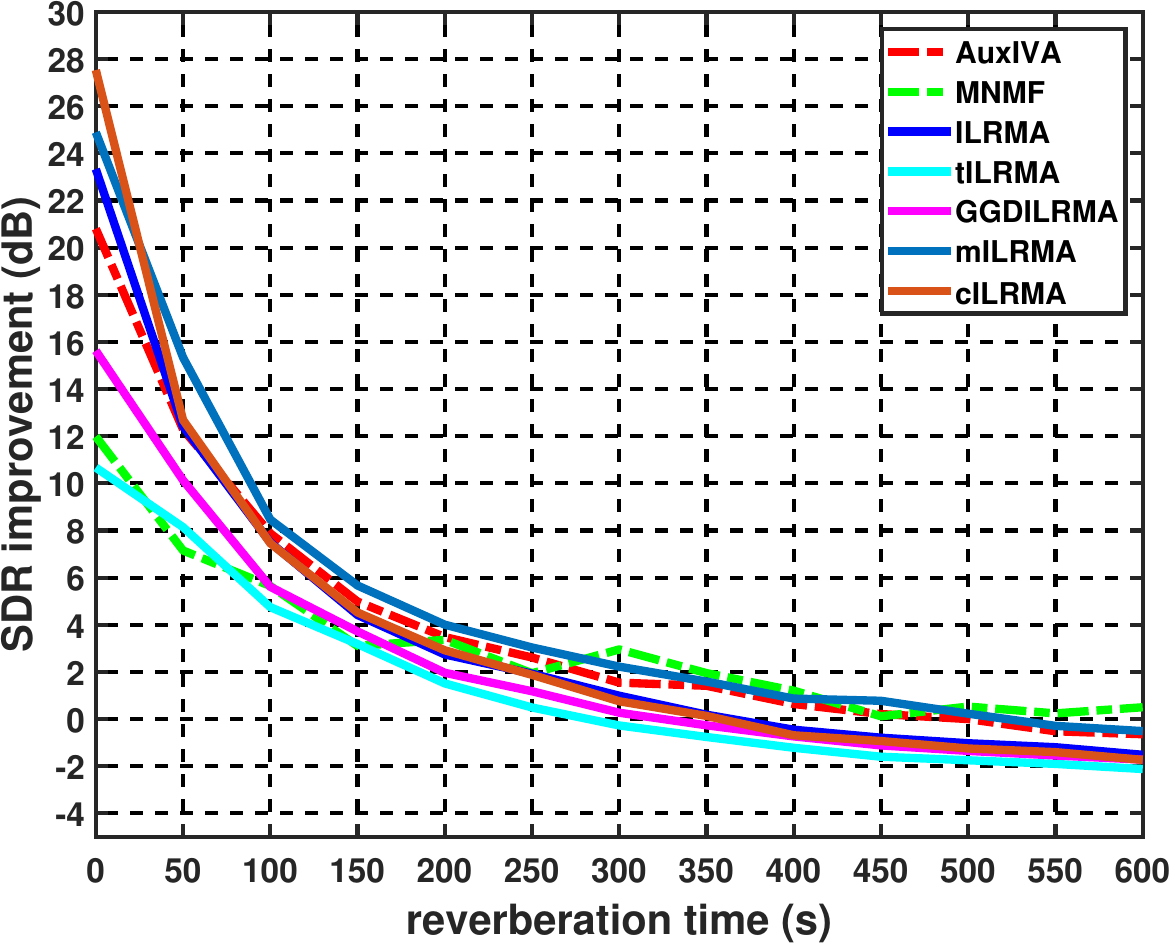}}
  \centerline{(e) female+male}\medskip
\end{minipage}
\begin{minipage}[b]{0.55\linewidth}
  \centering
  \centerline{\includegraphics[width=4.2cm]{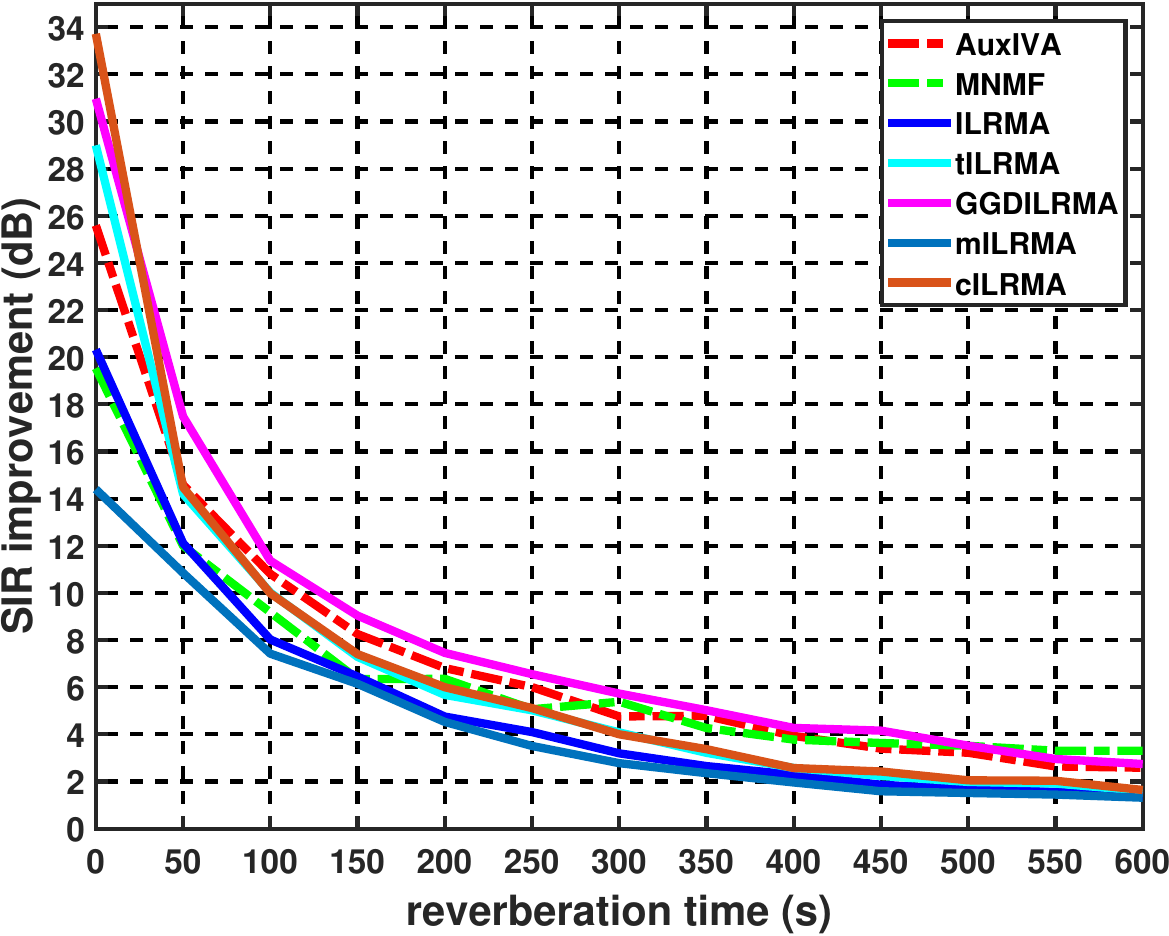}}
  \centerline{(f) female+male}\medskip
\end{minipage}
\vspace{-0.8cm}
\caption{SDR and SIR improvements for the studied methods.}
\label{fig:2}
\vspace{-0.5cm}
\end{figure}

\section{Experiments}
\subsection{Experimental configuration}
We followed the SISEC challenge \cite{araki20122011} and selected speech signals from
the Wall Street Journal (WSJ0) corpus \cite{garofolo1993csr} as the clean speech source
signals. Subsequently, we constructed evaluation signals for specific speech separation
task with $M=N=2$ and simulated room environments. The dimensions of the room were set
to $8\times 8\times 3$ m. Two microphones are positioned at the center of the room and their
spacing is $6$~cm. The two sources were positioned 2 m from the center of the two microphones.
The incident angles of the two source signals were designated as $80^\circ$ and $110^\circ$
respectively, with the direction normal to the line connecting the two microphones
marked as $0^\circ$. We employed the image model method \cite{allen1979image} to generate
room impulse responses where the sound absorption coefficients were determined using
Sabine's Formula \cite{young1959sabine}. The reverberation time $T_{60}$ is controlled
to be in the range from $0$ to $600$ ms with an interval of $50$ ms. For each gender
combinations (there are three combinations) and every value of $T_{60}$, 100 sets of mixed
signals are generated for evaluation. The sampling rate is $16$ kHz.

The value of the hyperparameter $\sigma$ in cILRMA was set to 1. We compared cILRMA with
AuxIVA \cite{ono2011stable}, MNMF \cite{sawada2013multichannel}, ILRMA \cite{kitamura2016determined},
$t$ILRMA \cite{mogami2017independent}, Generalized Gaussian distributed ILRMA (GGDILRMA) \cite{ikeshita2018independent}
and $m$ILRMA \cite{wang2021minimumap}. Signal-to-distortion ration (SDR) and source-to-interferences ratio (SIR) are
used as the performance metrics for evaluation and the definitions of these metrics can be found in \cite{vincent2006performance}.

\begin{figure}[t]
%
\begin{minipage}[b]{.46\linewidth}
  \centering
  \centerline{\includegraphics[width=4.2cm]{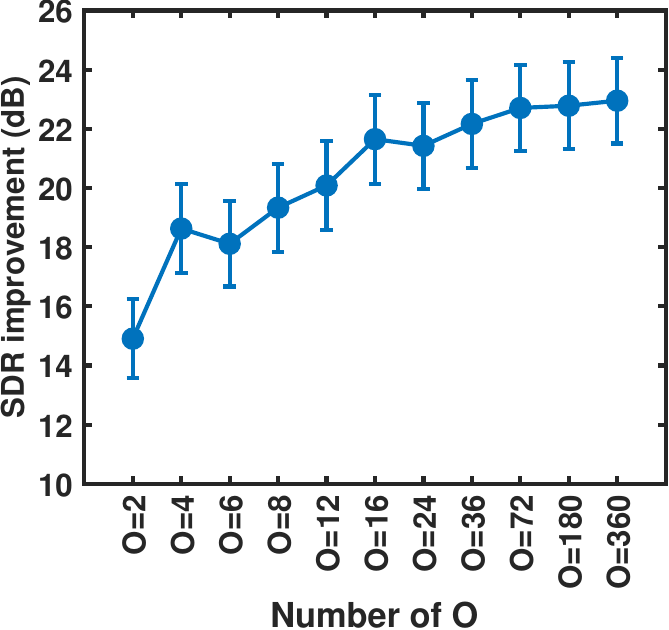}}
  \vspace{0.2cm}
\end{minipage}
\begin{minipage}[b]{.55\linewidth}
  \centering
  \centerline{\includegraphics[width=4.2cm]{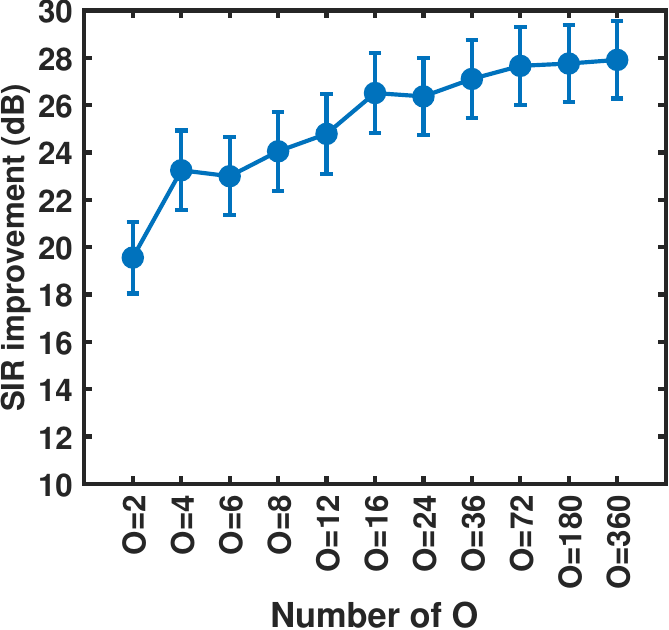}}
\vspace{0.2cm}
\end{minipage}
\vspace{-0.8cm}
\caption{Average SDR and SIR improvements versus the value of $O$. Conditions:
source signals are from two female speakers in WSJ0 and there is no reverberation.}
\label{fig:3}
\end{figure}

\begin{figure}[t]
%
\begin{minipage}[b]{.46\linewidth}
  \centering
  \centerline{\includegraphics[width=4.2cm]{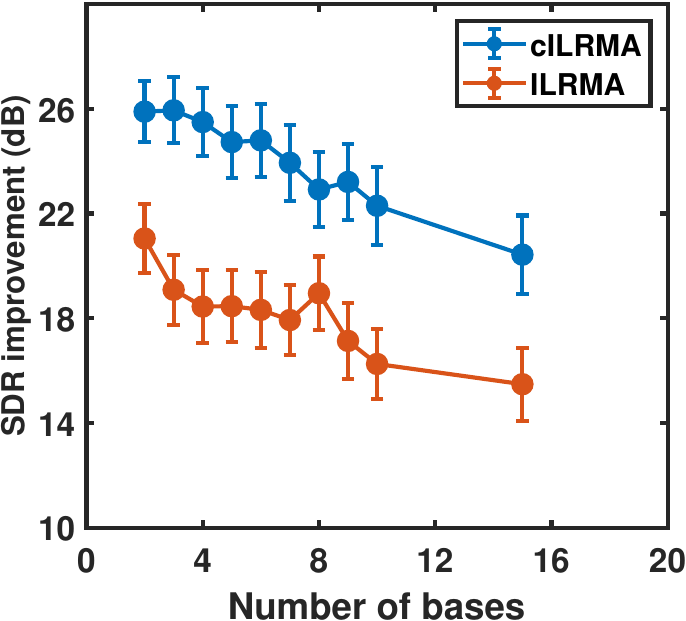}}
\vspace{0.2cm}
\end{minipage}
\begin{minipage}[b]{.55\linewidth}
  \centering
  \centerline{\includegraphics[width=4.2cm]{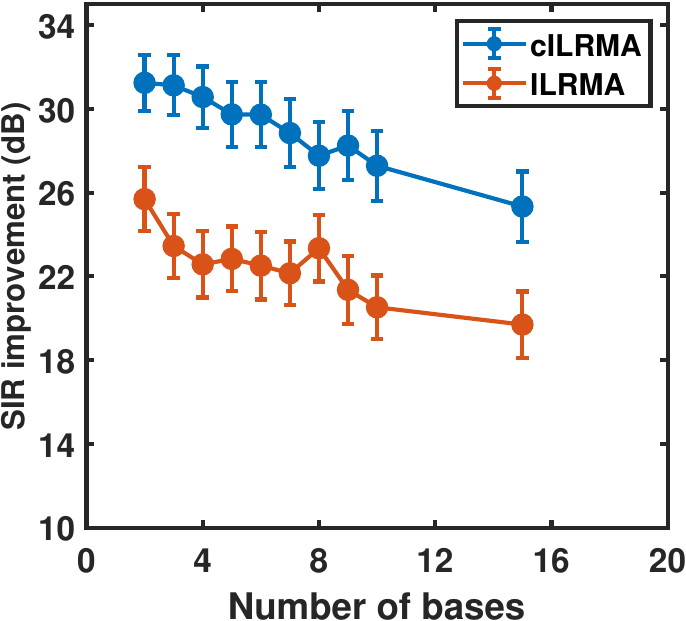}}
\vspace{0.2cm}
\end{minipage}
\vspace{-0.8cm}
\caption{Average SDR and SIR improvements versus different number of bases. Conditions:
source signals are from two female speakers in WSJ0 and there is no reverberation.}
\label{fig:4}
\end{figure}

\begin{figure}[!t]
%
\begin{minipage}[b]{.46\linewidth}
  \centering
  \centerline{\includegraphics[width=4.2cm]{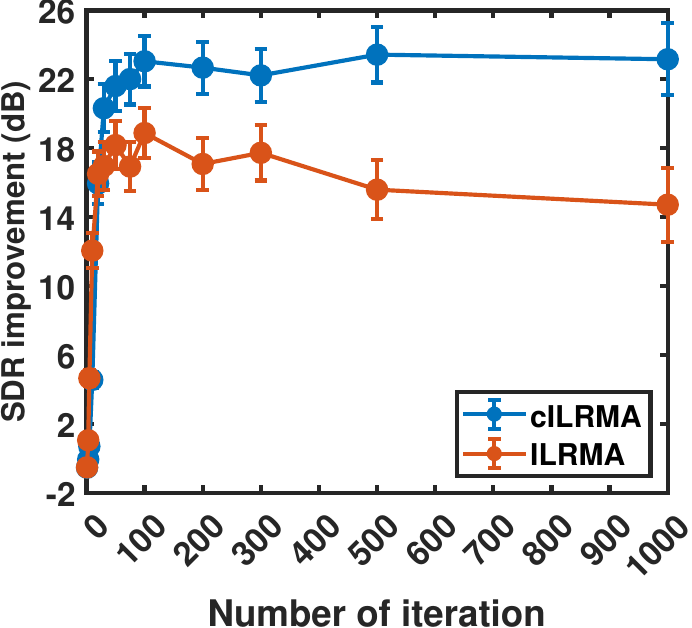}}
\vspace{0.2cm}
\end{minipage}
\begin{minipage}[b]{.55\linewidth}
  \centering
  \centerline{\includegraphics[width=4.2cm]{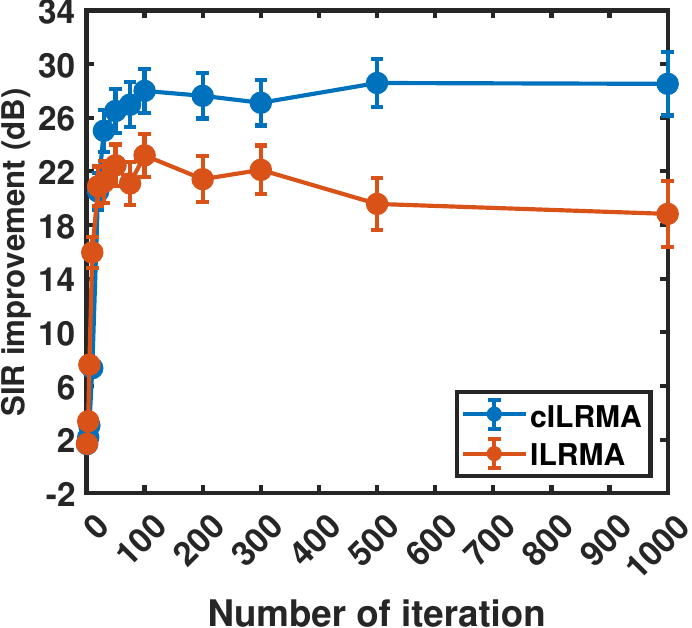}}
\vspace{0.2cm}
\end{minipage}
\vspace{-0.8cm}
\caption{Average SDR and SIR improvements versus iteration number.
Conditions: source signals are from two female speakers in WSJ0 and there is no reverberation.
}
\label{fig:5}
\vspace{-0.45cm}
\end{figure}

\subsection{Main results}

Figure~\ref{fig:2} plots the averaged performance in different reverberant environments.
It is evident that the cILRMA method exhibits greater improvements in SDR and SIR compared
to the other algorithms, although the discrepancy diminishes as the reverberation time increases.

An essential parameter in the proposed source model is parameter $O$. To explore its impact
on performance, a set of experiments were conducted. Figure~\ref{fig:3} depicts the
SDR and SIR improvements across various values of $O$, with the source signals originating
from two female speakers. The results indicate that performance enhances with increasing
values of $O$, suggesting that a higher $O$ value leads to a more accurate source model.

Figure~\ref{fig:4} illustrates the SDR and SIR improvements achieved by cILRMA and ILRMA
across varying numbers of bases. The results demonstrate that regardless of the number of
bases, cILRMA consistently outperforms ILRMA by approximately $4$ dB in terms of performance.

The convergence behavior for cILRMA and ILRMA are shown in Fig.~\ref{fig:5}.
It is seen that it is approximately 100 iterations for cILRMA to perform better than ILRMA.

\section{Conclusions}

The paper presented a clustered source model tailored for ILRMA-based
MBASS. Leveraging the NBTD technique, this model defines blocks as outer
products of vectors (clusters) and matrices for spectral structure modeling,
thus providing interpretable latent vectors. By integrating orthogonality
constraints, the model ensures independence among source images. Experimental
results demonstrated the superiority of the proposed method over its traditional
counterparts in anechoic environments.

\bibliographystyle{IEEEbib}

\end{document}